\documentclass[9pt,twocolumn,twoside]{opticajnl}
\journal{opticajournal} % use for journal or Optica Open submissions

\setboolean{shortarticle}{true}
\usepackage{stfloats}

\title{Determination of Plasma Properties in Liquid Jets Through Time-Resolved Experiments on Third Harmonic Reflection Dynamics}

\author[1,*]{S. Hilal}
\author[1]{M. Melnik}
\author[1]{A. Ismagilov}
\author[1]{A. Tsypkin}
\author[1]{S. Kozlov}

\affil[1]{Laboratory of Femtosecond and Femtotechnologies, ITMO University, Kadetskaya line., St. Petersburg, 197101, Russia}

\affil[*]{Shireenhilal@itmo.ru}

\begin{abstract}
The study of plasma in liquid jets represents a significant area of research encompassing plasma science, dynamics, and properties. This paper presents experimental studies on plasma formation processes in liquid jets of water, ethanol, and isopropyl based on the dynamics of the third harmonic reflection from the induced plasma surface. The characteristics of plasma formation and its properties concerning the angle of incidence of pump radiation on the jet are described. Through time-resolved experiments, and theoretical estimations using Keldysh theory, plasma properties, including reflectivity, density, and frequency for three different media: water, ethanol, and isopropyl are evaluated. Among the studied liquid jets, the isopropyl one demonstrates the highest values of the characteristics mentioned. These findings hold significant potential for advancing our understanding of plasma-based radiation sources e.g. terahertz generation.
\end{abstract}

\setboolean{displaycopyright}{false} % Do not include copyright or licensing information in submission.

\begin{document}

\maketitle

\section{Introduction}
Jets of liquids and gases are of great practical importance for various applications. Previously, they were used to create ultrashort laser systems \cite{YEOM2008499}. At the moment, they are widely used in X-ray \cite{mudryk2020electronic} and terahertz (THz) \cite{yiwen2018terahertz, tcypkin2019flat,ponomareva2021plasma} generation systems, as well as in laser-induced breakdown spectroscopy (LIBS) systems \cite{keerthi2022elemental,skovcovska2016optimization}. They are unique in the fact that, unlike stationary systems (cells or films), the jets constantly renew the area of interaction.

The effects associated with a breakdown in liquids and the study of plasma formation have been theoretically and experimentally analyzed in various works \cite{aghdam2020multiphysics,bruggeman2016plasma,vanraes2018plasma}. Plasma formation in liquids occurs at intensities lower than for gases, on the order of $10^{12}$ $W/cm^2$. Recently \cite{tcypkin2019flat} it was investigated how material parameters like linear absorption, molecular density and ionization energy impact the optical-to-THz conversion efficiency during THz generation via plasma formation in the medium. Furthermore, experiments revealed how the double-pulse excitation technique improves the conversion efficiency of THz generation up to 0.1\% in the $\alpha$-pinene liquid jet \cite{ponomareva2021varying}. Given the fundamental role that plasma formation in liquid jets plays in THz generation, further research in this area is necessary to enhance the efficiency of THz generation. 

Recently \cite{ponomareva2021plasma} the features of the behavior of plasma optical properties under different conditions of the experiment for flat jets of several liquids have been studied. An original technique for measuring plasma dynamics by means of the method of third harmonic (TH) reflection from induced plasma was proposed, which makes it possible to evaluate plasma properties. Nevertheless, in real systems, it is necessary to conduct research and identify certain patterns of plasma formation with respect to the nonlinear properties of the liquids used. Knowledge about the dynamics of plasma formation in liquid jets is necessary for a more detailed understanding of the generation of THz radiation \cite{ponomareva2019double, ponomareva2021varying}.

The first part of this work is aimed at studying the regularities of plasma properties with a change in the angle of incidence on the jet with respect to the normal. In the second part, we are implementing an experimental study on the plasma formation in isopropyl, ethanol, and water liquid jets based on the dynamics of TH reflection from the induced plasma. The results demonstrate the patterns of plasma formation when using various liquids which can help to choose medium for specific task. Theoretical analysis explains all the obtained experimental regularities. 

\begin{figure}[ht]
\centering\includegraphics[width=\linewidth]{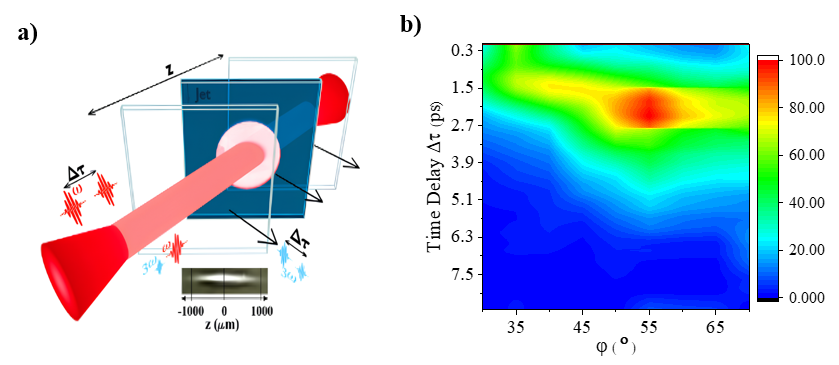}
\caption{\textbf{a}. Experimental scheme for studying the properties of a plasma formed during filamentation in a liquid jet by measuring the dynamics of plasma-induced reflection of the probe pulse radiation. The method of pump sample spectroscopy is used, provided that the first pulse forms a plasma in a liquid jet, and the second one, as a result of filamentation in front of the jet, generates TH in addition to the main one, which is also used as probe radiation. The time delay between pulses is controlled by the Michelson circuit. The registration of the reflected signal from the induced plasma by the first pulse is recorded using a spectrometer. \textbf{b}.The dependence of TH reflection dynamics on the jet rotation angle.}
\label{fig1}
\end{figure}

To study the plasma induced in liquid jets, an experimental setup (Fig.~\ref{fig1}a) was assembled. The method presented in this paper is a time-resolved approach utilizing the double pump technique based on \cite{ponomareva2021plasma}. As a result, the focusing system receives two femtosecond pulses at the input, shifted by a time delay between them $\Delta \tau$. The essence of the method is that the first pulse, forms a preionized medium, preparing it to reflect the tail of the same pulse \cite{ponomareva2021plasma}. It creates the conditions for the second pulse, shifted by time delay $\Delta\tau$,  to interact with the ionized region, increasing plasma surface reflectivity. The second pulse is a delayed copy of the first one so it carries the same energy and acts as a probe. When focusing both pulses onto the liquid jet's surface, the plasma induces in the liquid jet and TH generation occurs in the ionized air along the path of the laser pulses due to the high intensity\cite{ponomareva2021plasma}. When the incident radiation frequency is close to the plasma frequency, the plasma reflects the radiation. Therefore, employing the pump-probe method allows us to observe the reflection not only of the first harmonic but also of the TH and measure the reflection dynamics with an accuracy of up to tens of fs \cite{ponomareva2021plasma}. Notably, the studied TH signals are observed in two positions: one is in front of the liquid jet's surface at an angle not aligned with the pump pulses, and the second position is behind the liquid jet, aligning with the path of the pump pulses. However, no TH radiation was observed in the opposing direction relative to the detected TH signal in front of the liquid jet's surface. This confirms that the TH radiation is generated in the ionized air along the path of the pump pulses and reflected from the plasma formed in the liquid jet. This enables us to study the dynamics of TH reflection from the plasma formed in the liquid jet, ensuring that the reflected TH is uncorrelated with the properties of the liquid jet medium due to its reflection primarily from the induced plasma rather than direct interaction with the liquid jet. It is important to note that our study does not encompass the TH radiation originating from the second position located behind the liquid jet. This omission arises from the fact that the TH radiation detected in this position is not a result of reflection from the plasma formed within the liquid jet. Consequently, it does not contribute to our understanding of the plasma dynamics under investigation.

The liquid jet is located on a slider that can move it along the $z$ axis. It is known that when using a focal lens of $10$ $cm$, the waist length will be about $2$ $mm$. A photograph of a filament in the air without a jet using the same experimental conditions is shown in the inset in Fig.~\ref{fig1}(a). The jet is located at the central region of the filament ($z=0$). The radiation reflected from the plasma in the liquid jet is measured using a spectrometer. Each measurement is taken 40 times and averaged. The jet has a thickness of 100 $\mu m$ and is rotated for all measurements at a different angle concerning the incidence of pump radiation. The jet is formed using a nozzle \cite{watanabe1989new}. The pump radiation is a Gaussian beam generated by a femtosecond laser system with a central wavelength of 800 $nm$, a duration of 150 fs, FWHM waist diameter
is 100 $\mu m$, and pulse energy of 300 $\mu J$ for each pulse in the double pump pulses. In the case of two-pulse excitation of a liquid jet, the Michelson interferometer scheme is used and the time delay between pulses varies from 0 to 10 $ps$. The dynamics of the reflection of the TH are measured as a function of the incident pump radiation angle as shown in Fig.~\ref{fig1}(b). These measurements are performed using a liquid jet of water as the medium. The maximum reflected TH dynamics are observed within a time delay range of 0-3 ps, and within an angular range of 50-60$^{\circ}$, centered around the Brewster angle of water at approximately 53$^{\circ}$. Observations show that as the angle of incidence deviates from this optimal range, whether decreasing or increasing, the maxima exhibit a faster relaxation and minimal shift in position.

\begin{figure}[ht]
\centering
\includegraphics[width=\linewidth]{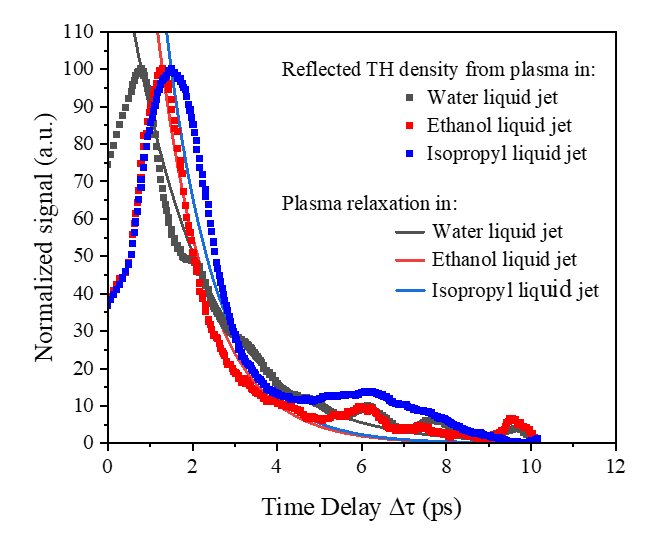}
\caption{Experimental results of TH (265 nm) density reflected from the induced plasma in water (black dots), ethanol (red dots), and isopropyl (blue dots) as a function of the time delay between the double pump pulses. With the exponential fitting curves of the experimental data describing the plasma relaxation in water (black line), ethanol (red line), and isopropyl (blue line), at a fixed energy of 300 $\mu$J and pulse duration of 150 fs.}
\label{fig2}
\end{figure}

Having chosen the conditions for the appropriate energy and pulse duration as outlined in \cite{ponomareva2021plasma}, we compare the properties of plasma formed in various liquids. The jet position is fixed at the center of the generated filament ($z = 0$) as shown in Fig.~\ref{fig1}(a), with a 55° rotation angle relative to the normal, corresponding to the average Brewster's angle for the studied liquids (water, ethanol, and isopropyl). The choice of liquids was due to their very different nonlinear characteristics. Fig.~\ref{fig2} shows the dynamics of the TH reflection from the induced plasma as a function of time delay between the double pump pulses at a fixed pulse energy of 300 $\mu$J and a fixed pulse duration of 150 fs.

Considering the interaction between the second pulse and the preionized region caused by the first pulse, the plasma density, and plasma frequency reach higher values for shorter time delays between the pump pulses.  Consequently, the plasma frequency initiates at its maximum, diminishing as the time delay increases. When the TH radiation frequency $\omega_{TH} = 7.06 \times 10^{15}$ Hz (256 nm) is much lower than the plasma frequency $\omega_p$, the plasma becomes absorptive, and the wave's energy transfers to the plasma, leading to attenuation. In this case, the attenuation of TH signal caused by the absorption, explains the low reflected TH density starting at zero time delay until the maximum. As time delay increases, the plasma frequency decreases and approaches values near $\omega_{TH}$, resulting in higher reflectivity. This explains the increasing in reflected TH density in Fig.~\ref{fig2} until reaching a maximum when $\omega_p \sim \omega_{TH}$.
 As the plasma frequency continues to decrease beyond this point, the reflectivity of the plasma diminishes. Eventually, when $\omega_p$ is smaller than $\omega_{TH}$, the plasma exhibits a weak response to the TH radiation due to the significantly higher energy associated with the wave compared to the characteristic energy of plasma oscillations. Consequently, the reflected density decreases which is seen in the figure.
To gain more insight, the plasma relaxation can be described by the measured TH reflection from the plasma as a function of the time delay between the double pump pulses by exponentially fitting the experimental curves in Fig.~\ref{fig2}, passing through the peaks, using the following exponential equation:
\begin{equation}
\omega_p(\Delta \tau)=\alpha \cdot e^{-\left(\Delta \tau-\tau_0\right) / \tau_d}
\end{equation}
 The parameter $\tau_d$ is a fitting parameter that represents the degradation time, which describes how quickly the plasma frequency is decreasing and, thus how fast the TH reflection occurs. The discrepancy between the experimental results for reflected TH density and the lines describing the plasma's relaxation behavior, for time delay values ranging from zero to those corresponding to the maximum peaks of TH densities, is due to the strong absorption of the plasma when $\omega_p > \omega_{TH}$. As a result, with delays shorter than those corresponding to the maximum, we have absorption.

\begin{figure}[ht]
\centering
\includegraphics[width=\linewidth]{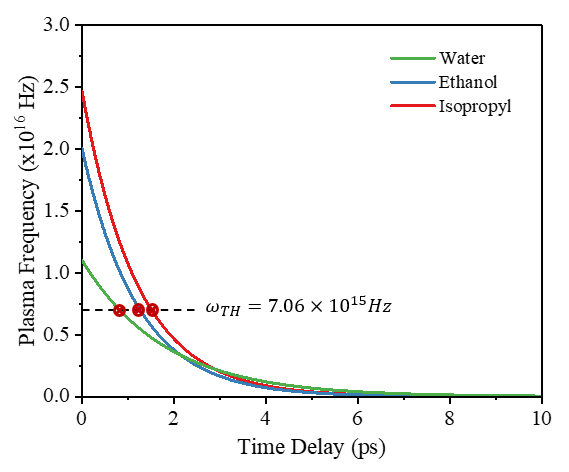}
\caption{Plasma frequency decreasing as a function of the time delay between the double pump pulses in the liquid jets of water (green line), ethanol (blue line), and isopropyl (red line). The red points correspond to the TH frequency at which the highest reflection occurs.}
\label{fig3}
\end{figure}

The chosen case clearly traces reflection dynamics changes with different liquids. Fitting the experimental curves with Eq. (1) reveals that plasma in water has a maximum degradation time of $\tau_d \sim 1.8$ ps, while ethanol and isopropyl show similar degradation times of approximately $\tau_d \sim 1.2$ ps. This difference can be attributed to their molecular structures and water's higher polarity. Ethanol and isopropyl's lower polarity leads to faster ionization processes, resulting in quicker changes in plasma frequency values. 
It can be seen from Fig.~\ref{fig2} that the maximum reflected TH density is observed for the liquid isopropyl at time delay $\Delta\tau$ $\sim$ 1.5 ps, and for ethanol liquid jet at lower time delay $\Delta\tau$ $\sim$ 1.25 ps, while the maximum reflection is observed for water liquid jet at the shortest time delay among the other two liquids $\Delta\tau$ $\sim$ 0.8 ps. As the maximum TH reflected density occurs when $\omega_p \sim \omega_{TH}$, the critical plasma frequency at the peak TH reflection corresponds to the TH frequency ($\omega_p \sim \omega_{TH} = 7.06 \times 10^{15} Hz$). By introducing this value of plasma frequency as $\alpha$ in Eq. (1) and associating degradation times $\tau_d$ with the peak position for each liquid, it becomes possible to estimate the plasma frequency as a function of the time delay $\Delta\tau$ between the double pump pulses as seen in Fig.~\ref{fig3}. 

In Fig.~\ref{fig3}, the plasma frequency values, obtained from the measured TH reflection dynamics of the plasma, are presented as a function of the time delay between the double pump pulses (from $\Delta\tau=0$ to $10$ ps). We estimated that the plasma frequency decreases rapidly in water liquid jets (from $1.1 \times 10^{16}$ Hz to $0.42 \times 10^{14}$ Hz) and in ethanol liquid jets (from $2.00 \times 10^{16}$ Hz to $0.48 \times 10^{13}$ Hz), while in isopropyl liquid jets (from $2.46 \times 10^{16}$ Hz to $0.56 \times 10^{14}$ Hz). It is now feasible to estimate the variation in the maximum plasma frequency values among the studied liquid jets at zero time delay, where the plasma is at its highest frequency. The isopropyl liquid jet shows a maximum plasma frequency 2.24 times higher than that of water, while ethanol demonstrates a maximum plasma frequency 1.8 times greater than water. The effects of plasma formation are known to be the main cause of generating THz fields \cite{ponomareva2021varying, ismagilov2021liquid}. Our findings align closely with the efficiency ratio of THz generation in these liquids, where the THz generation efficiency of isopropyl is twice that of water, while ethanol's efficiency surpasses that of water by 1.25 times \cite{ismagilov2021liquid} confirming the significant role of plasma.

In order to interpret the experimental data, Keldysh theory is considered \cite{Keldysh1965ionization,gruzdev2004analysis, gruzdev2005laser,bauer2016keldysh}, the theory of prediction of the multiphoton atomic ionization regimes under a strong laser field. Given that the complete ionization model is taken into consideration, with diffusion and recombination mechanisms ignored since they are not noticeable on the sub-picosecond timescale, then according to Keldysh theory, the induced plasma electron density is determined by considering both tunneling and multiphoton ionization phenomena. This density is calculated by solving the ionization rate equation  $\frac{\partial \rho}{\partial t}= W_{NP}$ where $W_{NP}$ (Eq.(37) in \cite{Keldysh1965ionization}):
\begin{multline}
W_{NP} =\frac{2 \omega}{9 \pi} \cdot\left(\frac{m_e \omega}{\gamma_1 \hbar}\right)^{3/2}  \cdot Q(\gamma, x)\\ 
\cdot\exp \left(-\pi\langle x+1\rangle \cdot\frac{K\left(\gamma_1\right)-E\left(\gamma_1\right)}{E\left(\gamma_2\right)}\right)
\end{multline}
Where $\rho$ is the plasma electron density, and the plasma frequency is given as $\omega_p = \sqrt{\rho e^2 /m_e \varepsilon_0}$, with Keldysh parameter given by the formula:
\begin{equation}
\gamma=\frac{\omega\sqrt{m_e \Delta}}{eF} 
\end{equation}
Where \begin{math}\gamma_1={\gamma}/{\sqrt{1+\gamma^2}}\end{math} and \begin{math}\gamma_2={\gamma_1}/{\gamma}\end{math}, $\omega$ is the angular frequency of the laser field,  $e$ is the electron charge, $m_e$ is the electron mass, $\varepsilon_0$ is the vacuum permittivity,  $\Delta$ is the material bandgap,  $F$ is the laser electric field amplitude, $\hbar$ is the reduced Planck constant, $K$ and $E$ are the complete elliptic integrals functions of the first and second kind respectively, and $Q(\gamma, x)$ is a function pertains to the structure of the spectrum, which is associated with the discreteness of the absorbed photon count, given by the corrected formula \cite{gruzdev2005laser}: 
\begin{multline}
Q(\gamma, x)= \left(\sqrt{\frac{\pi}{2 K\left(\gamma_2\right)}}\right) \sum_{n=0}^{\infty}\left[\exp \left(-\pi n \cdot \frac{K\left(\gamma_1\right)- E\left(\gamma_1\right)}{E\left(\gamma_2\right)}\right) \right.\\ \left.\cdot \Phi\left(\sqrt{\frac{\pi^2(\langle x+1\rangle-x+n)}{2 K\left(\gamma_2\right)E\left(\gamma_2\right)}}\right)\right]
\end{multline}
Where $n$ is the number of absorbed photons in the multiphoton absorption effect and the function $\Phi$ can be defined by $\Phi(z)=\int_0^z e^{y^2-z^2} d y$ and $x={\Delta^*}/{\hbar\omega}$.

The key medium parameters are the bandgap energy ($\Delta$) and the refractive index. The medium under the laser field can be described using the effective ionization potential formula:
\begin{equation}
\Delta^*=\frac{2\Delta E(\gamma_2)}{\pi \gamma_1}
\end{equation}
 Given that the concept of bandgap energy pertains to solid-state physics, it is more appropriate to prioritize the consideration of ionization energy rather than bandgap energy when discussing liquid mediums \cite{sollier2001numerical}. The approximate ionization energy values for Ethanol, Water, and Isopropyl are as follows: 10.7 eV \cite{von198030}, 11.67 eV \cite{perry2020ionization}, and 10.49 eV \cite{peel1975photoelectron}, respectively. 
With the provided pump parameters, it becomes feasible to estimate the plasma density by obtaining $W_{NP}$ and integrating over the pump pulse duration \cite{kennedy2003first}, thereby calculating the plasma frequency in each liquid jet.
Keldysh theory allows us to estimate the plasma parameters in a medium excited by a single pulse of the double pump pulse technique.  Following ionization by the initial pulse after a time delay $\Delta\tau$, the determination of crucial medium parameters such as ionization energy and refractive index becomes unattainable. Consequently, the introduction of a second pulse after time delay $\Delta\tau$ into the calculation is precluded.
However, In order to compare with the predicted maximum plasma frequency values obtained in Fig.~\ref{fig3}, it is possible to estimate the effect of double pulse excitation when the time delay between them is zero $\Delta\tau = 0$. In this case, the liquid jet is simultaneously pumped by the two pulses, resulting in the liquid jet surface receiving double the intensity of the single pulse in the double pump configuration. Table ~\ref{tab:my_label} presents the calculated plasma density and frequency based on a single pulse and based on double pulses when $\Delta\tau = 0$.

\begin{table*}[bp]
  \centering
  \caption{\bf The calculated values of  plasma  densities and frequencies formed in the studied liquid jets based on a single pulse and based on double pulses when $\Delta\tau = 0$, with the estimated plasma frequency obtained experimentally in Fig.~\ref{fig3}}
    \label{tab:my_label}
    \begin{tabular}{c c c c c c}
    \hline
  & \multicolumn{2}{c}{Single-Pulse} & \multicolumn{3}{c}{Double-Pulse $\Delta\tau = 0$}\\
     \hline
 Liquid Jet&$\begin{array}{c}
\text {Plasma Density}\\ (cm^{-3})
\end{array}$ & $\begin{array}{c}
\text {Plasma Frequency} \\ {(Hz)}\end{array}$ & $\begin{array}{c}
\text {Plasma Density}\\ (cm^{-3})\end{array}$ & $\begin{array}{c}
\text {Plasma Frequency} \\ {(Hz)}\end{array}$ & $\begin{array}{c}\text {Plasma frequency} \\ \text {from Fig.~\ref{fig3} (Hz)}\end{array}$ \\
\hline Water & $2.99 \times 10^{21}$ & $3.08 \times 10^{15}$ & $5.57 \times 10^{22}$ & $1.33 \times 10^{16}$ & $1.1 \times 10^{16}$\\
Ethanol &  $1.16 \times 10^{22}$ & $6.07 \times 10^{15}$ & $1.57 \times 10^{23}$ & $2.23 \times 10^{16}$ & $2.01 \times 10^{16}$\\
Isopropyl & $1.40 \times 10^{22}$ & $6.69\times 10^{15}$ & $1.92 \times 10^{23}$ & $2.47 \times 10^{16}$ & $2.46 \times 10^{16}$
\\
\hline
\end{tabular}
\end{table*}

The calculated plasma frequency values based on the single pulse are initially lower than the TH frequency $\omega_{TH} = 7.06 \times 10^{15} Hz$ ($256$ $nm$), indicating that the first pulse alone is insufficient to achieve maximum TH reflected density. However with introducing the second pulse of the double pump pulse technique, ionization is enhanced, resulting in plasma frequency values greater than the TH frequency as evident in the calculations for the double pulse at $\Delta\tau = 0$.  This observation aligns with previous studies \cite{sattmann1995laser,st1998analysis} that highlight the augmentation of ionization when employing the double pump pulse technique.
On the other hand, the calculated plasma frequency values based on double pump pulse excitation at $\Delta\tau = 0$ closely align with the experimentally estimated values at $\Delta\tau = 0$ from Fig.~\ref{fig3}, showing slightly higher values in the theoretical model. This validates the accuracy of our experimental estimation and explanation.

In conclusion, our study investigated plasma formation in liquid jets, focusing on TH reflection dynamics from induced plasma. We have examined experimentally how the jet rotation angle influenced TH reflection, observing maximum reflection at the Brewster angle for the liquid jet used.
For plasma property evaluation, we have determined the plasma frequency values from time-resolved experiment on the dynamics of TH reflection. We have compared water, ethanol, and isopropyl liquid jets. Isopropyl has exhibited the highest plasma frequency, ethanol lower, and water the lowest, correlating with their terahertz (THz) generation efficiency \cite{ismagilov2021liquid}. Utilizing Keldysh's theory, we have employed a theoretical model that estimated plasma parameters, aligning well with experimental evaluations in the investigated liquid jets. Furthermore, we have shown that using the double pump technique enhances the ionization in liquids resulting in increasing the plasma density and frequency. Our results regarding the variation of plasma frequency among the studied liquids are consistent with the efficiency of THz generation in these liquids, which is significantly influenced by the process of plasma formation \cite{ponomareva2021varying, ismagilov2021liquid}. These findings hold significant potential for the development of plasma as a THz radiation source.

\begin{backmatter}
\bmsection{Funding} This work was supported by the Russian Science Foundation (RSF) (2019-0903).

\bmsection{Disclosures} The authors declare no conflicts of interests.

\bmsection{Data Availability Statement} Data underlying the results presented in this paper are not publicly available at this time but may be obtained from the authors upon reasonable request.
\end{backmatter}
% Bibliography
\bibliography{sample}
\bibliographyfullrefs{sample}
\end{document}